%% file: manuscript.tex
\documentclass[aps,prl,twocolumn,superscriptaddress]{revtex4-2}

\usepackage{chngcntr}
\usepackage{graphicx}         
\usepackage{bm}               
\usepackage{amssymb}          
\usepackage{amsmath}          
\usepackage{CJK}              
\usepackage[dvipsnames]{xcolor} 
\usepackage{orcidlink}

\newcommand {\eV}          {\,\rm eV}

\newcommand {\kms}         {\,\rm km\,s^{-1}}

\newcommand {\kpc}         {\,\rm kpc}
\newcommand {\Mpc}         {\,\rm Mpc}

\newcommand {\Msun}        {\,\rm{M}_{\odot}}
\newcommand {\Ms}          {M_{\rm s}}
\newcommand {\Mh}          {M_{\rm h}}
\newcommand {\rh}          {r_{\rm h}}

\newcommand {\avewh}       {\ave{w}_{\rm h}}

\newcommand {\avevh}       {\ave{v}_{\rm h}}
\newcommand {\rs}          {r_{\rm s}}

\newcommand {\avews}       {\ave{w}_{\rm s}}

\newcommand {\whin}        {w_{\rm h, in}}
\newcommand {\vhin}        {v_{\rm h, in}}
\newcommand {\mFDM}        {m_{\rm 22}}
\newcommand {\LdB}         {\lambda_{\rm dB}}
\newcommand {\Mssim}       {M_{\rm s, sim}}
\newcommand {\Msthe}       {M_{\rm s, theory}}
\newcommand {\Mhalf}       {M_{\rm 1/2}}
\newcommand {\Ep}          { E_{\rm p} }
\newcommand {\Ek}          { E_{\rm k} }

\newcommand {\fref}[1]     {Fig.~\ref{#1}}
\newcommand {\tref}[1]     {Table~\ref{#1}}
\newcommand {\eref}[1]     {Eq.~(\ref{#1})}

\newcommand {\be}          {\begin{equation}}
\newcommand {\ee}          {\end{equation}}
\newcommand {\ave}[1]      {\langle #1 \rangle}

\begin{document}
\begin{CJK*}{UTF8}{bkai}

\title{Deciphering the Soliton-Halo Relation in Fuzzy Dark Matter}

\author{Pin-Yu Liao (廖品瑜)\orcidlink{0009-0007-8469-5880}}
\affiliation{Institute of Astrophysics, National Taiwan University, Taipei 10617, Taiwan}
\affiliation{Department of Physics, National Taiwan University, Taipei 10617, Taiwan}

\author{Guan-Ming Su (蘇冠銘)\orcidlink{0009-0008-4889-9779}}
\affiliation{Department of Physics, National Taiwan University, Taipei 10617, Taiwan}

\author{Hsi-Yu Schive (薛熙于)\orcidlink{0000-0002-1249-279X}}
\email{hyschive@phys.ntu.edu.tw}
\affiliation{Institute of Astrophysics, National Taiwan University, Taipei 10617, Taiwan}
\affiliation{Department of Physics, National Taiwan University, Taipei 10617, Taiwan}
\affiliation{Center for Theoretical Physics, National Taiwan University, Taipei 10617, Taiwan}
\affiliation{Physics Division, National Center for Theoretical Sciences, Taipei 10617, Taiwan}

\author{Alexander Kunkel\orcidlink{0009-0007-2957-4277}}
\affiliation{Institute of Astrophysics, National Taiwan University, Taipei 10617, Taiwan}
\affiliation{Department of Physics, National Taiwan University, Taipei 10617, Taiwan}

\author{Hsinhao Huang (黃新豪)\orcidlink{/0000-0002-7368-1324}}
\affiliation{Institute of Astrophysics, National Taiwan University, Taipei 10617, Taiwan}
\affiliation{Department of Physics, National Taiwan University, Taipei 10617, Taiwan}

\author{Tzihong Chiueh (闕志鴻)\orcidlink{0000-0003-2654-8763}}
\affiliation{Institute of Astrophysics, National Taiwan University, Taipei 10617, Taiwan}
\affiliation{Department of Physics, National Taiwan University, Taipei 10617, Taiwan}
\affiliation{Center for Theoretical Physics, National Taiwan University, Taipei 10617, Taiwan}

\date{\today}

\begin{abstract}
Soliton cores at the center of fuzzy dark matter (FDM) halos provide a promising way to distinguish FDM from other dark matter models. However, the relation between solitons and their host halos remains contentious. Here, we rigorously examine this soliton-halo relation (SHR) using a rich set of cosmological simulations across various FDM particle masses, halo masses, and redshifts. We explicitly demonstrate thermal equilibrium between solitons and surrounding halo granules, energy equipartition within halos, and an FDM concentration-mass-nonisothermality relation. For each FDM halo, we confirm that its density profile outside the central soliton matches a collisionless N-body simulation from the same initial condition, serving as stringent numerical convergence tests. Our refined SHR agrees well with virialized halos in simulations, with a $1\sigma$ deviation of less than $30\%$. These findings not only reaffirm the SHR proposed by Schive et al. (2014) but also offer a more comprehensive understanding that extends its applicability.
\end{abstract}

\maketitle
\end{CJK*}

\textit{Introduction}.---Fuzzy dark matter (FDM) \cite{hu2000, peebles2000, Guzman2000, goodman2000, Bohmer2007, Sikivie2009, Woo2009, marsh2016, hui2017, hui_wave_2021, ChadhaDay2022}, composed of ultra-light bosons with masses around $m\sim10^{-22}$--$10^{-20}\eV$, offers a compelling alternative to cold dark matter (CDM). Its sub-kiloparsec de Broglie wavelength produces distinctive quantum wave effects on galactic scales, including the suppression of low-mass halos \cite{Schive2016, may2021, May2023}, the formation of dynamic, granular structures throughout halos \cite{schive2014a, veltmaat2018, yang2024}, and the presence of stable soliton cores surrounded by a Navarro-Frenk-White (NFW; \cite{NFW}) profile \cite{schive2014a, schive2014b, mocz2017, marsh2015, Khlopov1985}. Solitons are spherically symmetric, stationary, ground-state solutions of the Schr\"{o}dinger-Poisson equation \cite{Ruffini1969}, where quantum pressure counteracts gravity, producing a dense, flat central density profile. Additionally, solitons exhibit density oscillations and random walk due to wave interference \cite{veltmaat2018, schive2020, li2021, Chiang2021, Chowdhury2021, zagorac2022}. These characteristics are markedly different from the predictions of CDM and other alternative dark matter models, thereby providing a promising means of discriminating FDM. It is therefore critical to quantitatively examine how soliton properties depend on their host halos --- the soliton-halo relation (SHR).

For a fixed $m$, all soliton solutions satisfy a scale transformation \cite{Ruffini1969, Guzman2006} characterized by a single parameter, such as the soliton mass $\Ms$. Therefore, the SHR is typically expressed as $\Ms(\Mh,z,m)$, where $\Mh$ is the host halo mass and $z$ is redshift. Schive et al. \cite{schive2014b} proposed $\Ms \propto m^{-1}(1+z)^{1/2}\Mh^{1/3}$, derived from both cosmological simulations and a spherical top-hat collapse model assuming thermal equilibrium between the soliton and halo. This relation has been confirmed by several follow-up studies, including the cosmological simulations of \cite{veltmaat2018, may2021} and theoretical models based on halo mergers \cite{du2017} and thermodynamic approaches \cite{chavanis2019}, with further refinements incorporating NFW halos \cite{bar2018}, FDM concentration-halo relation \cite{taruya2022, kawai2024}, and self-interaction \cite{padilla2021}. However, several controversies exist. For example, the theoretical models of \cite{schive2014b, taruya2022, kawai2024} tend to underestimate $\Ms$ for massive halos. Mocz et al. \cite{mocz2017} proposed a different SHR, $\Ms \propto \Mh^{5/9}$, aligning with the simulation results of \cite{nori2021, mina2022}. Large scatter in $\Ms$ has been reported in \cite{schwabe2016, chan2022}, which, together with the aforementioned discrepancies, suggests that a universal SHR may not exist \cite{yavetz2022, zagorac2023, Dmitriev2024}. Moreover, the SHR and the soliton radius-mass scaling at fixed $m$ are in tension with observational constraints from galactic rotation curves \cite{Bar2022, Khelashvili2023, BanaresHernandez2023}, possibly alleviated by non-negligible self-interactions \cite{Delgado2023, Dave2023, Indjin2025}. These controversies necessitate further investigation to gain a deeper understanding of this relation.

Furthermore, FDM simulations are extremely challenging. Simulations solving the Schr\"{o}dinger-Poisson equation (e.g., \cite{schive2014b, mocz2017}) require much higher resolution than CDM simulations to resolve the de Broglie wavelength associated with high-speed flows. On the other hand, it remains unclear whether simulations based on the fluid-like Madelung equation (e.g., \cite{nori2021}) can accurately capture solitons, as they struggle to handle strong destructive interference with vanishing density that is ubiquitous in FDM halos \cite{Li2019}. Consequently, previous simulations addressing the SHR have been subject to significant uncertainties.

\begin{figure}[t!]
\includegraphics[width=\columnwidth]{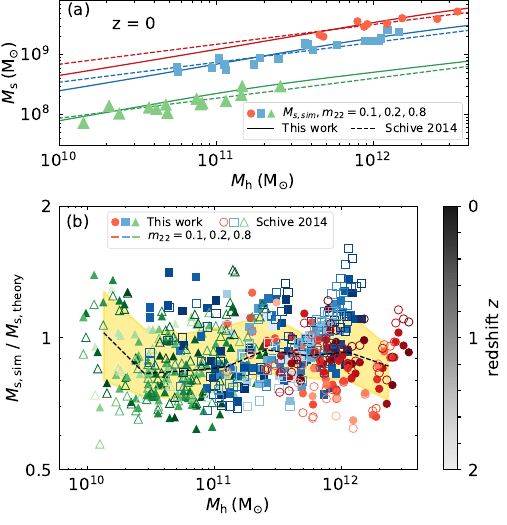}
\caption{
Comparison of soliton masses from cosmological simulations $\Mssim$ with theoretical predictions $\Msthe$ from Schive et al. (2014) \cite{schive2014b} and this work, as a function of halo mass $\Mh$. Different colors represent different FDM particle mass $\mFDM$. (a) $\Mssim$ (closed symbols) and $\Msthe$ (solid and dashed lines) at $z=0$. Halos in smaller-$\mFDM$ simulations are systematically more massive due to larger simulation boxes and the stronger suppression of low-mass halos. (b) $\Mssim/\Msthe$ at $z=0$--$2$. The dashed line shows the median values and the shaded region denotes the $1\sigma$ uncertainty in this work. Both theoretical models show good agreement with simulations over a wide parameter range: $\mFDM=0.1$--$0.8$, $\Mh \sim 7\times10^{9}$--$3\times10^{12}\Msun$, and $z=0$--$2$. Our work, however, provides significantly deeper insights into the soliton-halo relation (see Figs. \ref{fig:profile}--\ref{fig:ratios} and text for details).
}
\label{fig:sim_vs_theory}
\end{figure}

\begin{figure}[t!]
\includegraphics[width=\columnwidth]{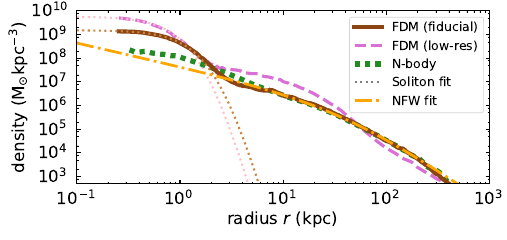}
\caption{
Density profiles of a representative FDM halo with $\Mh=9.8\times 10^{11} \Msun$, $\Ms = 1.7\times 10^{9}\Msun$, and $\mFDM=0.2$ at $z=0$. The fiducial result (solid line) closely matches both the N-body simulation (thick dotted line) and the NFW model (dash-dotted line) in the halo profile at $r \gtrsim 7\kpc$. The central profile fits well with the analytical soliton solution (thin dotted line). By contrast, the low-resolution result (dashed line) produces an overly concentrated halo and a soliton that is $30\%$ more massive, despite that the central profile still agrees with the analytical soliton solution. This discrepancy arises from an underestimation of quantum pressure, highlighting the importance of validating numerical convergence in FDM simulations.
}
\label{fig:profile}
\end{figure}

\begin{figure}[t!]
\includegraphics[width=\columnwidth]{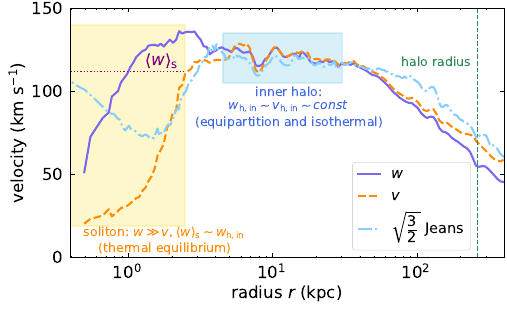}
\caption{
Velocity profiles of the same halo shown in \fref{fig:profile}. The solid and dashed lines represent the thermal velocity $w$ and bulk velocity $v$, respectively. In the inner halo surrounding the soliton (blue shaded region), the thermal velocity $\whin$ and bulk velocity $\vhin$ have similar magnitudes and remain roughly constant, indicating energy equipartition and an isothermal distribution. Both velocities closely match the velocity dispersion derived from the isotropic spherical Jeans equation (dash-dotted line). In contrast, the outer halo exhibits a non-isothermal velocity profile (see panel (b) of \fref{fig:conc_and_temp} for the FDM nonisothermality-concentration relation).
Within the soliton (yellow shaded region), the thermal velocity dominates, and its average value $\avews$ approximately coincides with $\whin$, suggesting thermal equilibrium between the soliton and inner halo. See panels (b) and (c) of \fref{fig:ratios} for an analysis of energy equipartition and thermal equilibrium across all halos.
}
\label{fig:velocity}
\end{figure}

\begin{figure}[t!]
\includegraphics[width=\columnwidth]{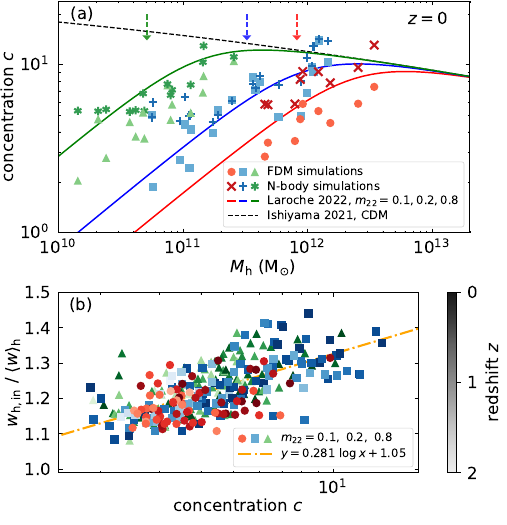}
\caption{(a) FDM concentration-mass relation at $z=0$. Different colors represent different $\mFDM$. Compared to CDM (dashed line), FDM halo concentration decreases for $\Mh$ below a half-mode halo mass, $\Mhalf \propto \mFDM^{-4/3}$ \cite{Schive2016} (vertical arrows), consistent with the theoretical predictions of \cite{laroche2022,kawai2024} (solid lines). For reference, we also include the results of N-body simulations using the same FDM initial conditions.
(b) FDM nonisothermality-concentration relation. Nonisothermality is defined as the ratio of the inner-halo thermal velocity, $\whin$, to the average thermal velocity of the entire halo, $\avewh$ (i.e., $\beta$ in \eref{eq:Ws}), which increases with higher concentration parameters. The dash-dotted line represents the regression fit. Notably, this relation is insensitive to redshift.
}
\label{fig:conc_and_temp}
\end{figure}

\begin{figure}[t!]
\includegraphics[width=\columnwidth]{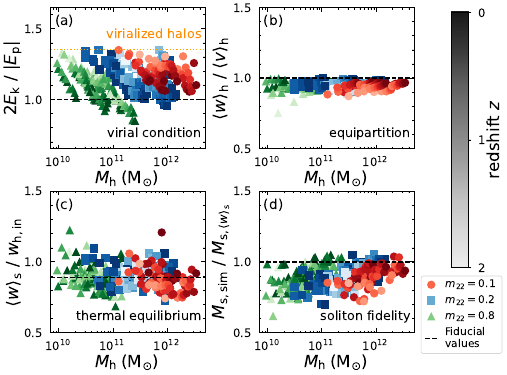}
\caption{
Examination of the individual terms and assumptions in the soliton-halo relation, Eqs. (\ref{eq:Ms}) and (\ref{eq:Ws}), for all FDM halos in our simulations. (a) Virial ratio $2\Ek/|\Ep|$, where $\Ek$ is the sum of thermal and kinetic energy and $\Ep$ is the potential energy. In this work, we include only virialized halos with $2\Ek/|\Ep| < 1.35$ (dotted line). (b) Ratio of the average thermal velocity $\avewh$ to the average bulk velocity $\avevh$ for the entire halo, equal to unity under energy equipartition. (c) Ratio of the average thermal velocity of the soliton $\avews$ to the thermal velocity in the inner halo surrounding the soliton $\whin$ (i.e., $\gamma$ in \eref{eq:Ws}), which is unity assuming thermal equilibrium. (d) Ratio of the soliton mass measured directly from the simulated density field, $M_{\rm s, sim}$, to that inferred from $\avews$, $M_{{\rm s}, \avews}$. It equals unity if a simulated central density profile follows exactly the analytical soliton solution. Dashed lines represent the fiducial values adopted in this work: $2\Ek/|\Ep|=\avewh/\avevh=M_{\rm s, sim}/M_{{\rm s}, \avews}=1$ and $\avews/\whin=0.89$.  
}
\label{fig:ratios}
\end{figure}

In this Letter, we propose an improved SHR, inferred from a rich set of cosmological simulations (\fref{fig:sim_vs_theory}) with rigorous convergence tests (\fref{fig:profile}). Our analysis addresses several key aspects that were previously unexplored, including the velocity distribution of FDM halos and solitons (\fref{fig:velocity}), the relations between FDM concentration, halo mass, and nonisothermality (\fref{fig:conc_and_temp}), and the validation of virialization, thermal equilibrium, energy equipartition, and soliton fidelity (\fref{fig:ratios}). We emphasize that although the small FDM particle masses explored in our simulations, $\mFDM \equiv m/10^{-22}\eV=0.1$--$0.8$, are disfavored by some astrophysical constraints \cite{Marsh2019, Rogers2021, Nadler2021, Dalal2022}, the inferred SHR is general and applicable to larger $\mFDM$. We describe simulation details in the \emph{Supplemental Material} \cite{SM}.

\textit{Simulation setup}.---We conduct FDM cosmological simulations using the GPU-accelerated adaptive mesh refinement code \texttt{GAMER} \cite{GAMER}. Notably, the code employs a novel hybrid integration method \cite{Kunkel2024}: on coarser grids, it uses a fluid approach to solve the Hamilton-Jacobi-Madelung equations, effectively capturing the large-scale structure without the need to resolve the de Broglie wavelength $\LdB$; on finer grids, it switches to a local pseudospectral scheme based on Fourier continuations with Gram polynomials to solve the Schr\"{o}dinger equation, thereby resolving small-scale interference fringes and solitons. Initial conditions are generated at $z=100$ using \texttt{AxionCAMB} \cite{AxionCAMB} and \texttt{MUSIC} \cite{MUSIC}. To investigate the $m$, $\Mh$, and $z$ dependence in the SHR, we perform 31 simulations with $\mFDM=(0.1, 0.2, 0.8)$. These simulations produce a total of 42 halos at $z=0$ without mergers, spanning a halo mass range of $\Mh\sim7\times10^{9}$--$3\times10^{12}\Msun$ at $z=0$--$2$.

To ensure the robustness of our FDM simulations, we increase the simulation resolution until the halo properties converge. Furthermore, for each FDM simulation, we conduct a collisionless N-body simulation with the same initial condition using the code \texttt{GADGET-2} \cite{Gadget2} to verify that the FDM and N-body halo density profiles match well outside the central soliton. See the \textit{Numerical convergence} section and \emph{Supplemental Material} \cite{SM} for details.

\textit{Soliton-halo relation}.---The improved SHR proposed in this work stems from the thermal and bulk velocity distributions of halos and solitons:
\be
\Ms(\Mh,z,\mFDM) = 3.23 \times 10^8 \left( \frac{\avews}{100\kms} \right) \mFDM^{-1} \Msun,
\label{eq:Ms}
\ee
\be
\avews = \left( \frac{|\Ep(\Mh,c)|}{\Mh} \right)^{1/2} \alpha \beta(c) \gamma.
\label{eq:Ws}
\ee
Here $\Ep(\Mh,c)$ is the gravitational potential energy and $c(\Mh,z,\mFDM)$ is the concentration parameter assuming an NFW halo (see the \textit{Concentration-mass-nonisothermality relation} section for details). We define $\alpha \equiv ( 1 + (\avevh/\avewh)^2 )^{-1/2}$, $\beta \equiv \whin/\avewh$, and $\gamma \equiv \avews/\whin$. $\bm{w}=(\hbar/2m)\bm{\nabla}\rho/\rho$ is the thermal velocity and $\bm{v}=(\hbar/m)\bm{\nabla} S$ is the bulk velocity, with the wave function expressed as $\psi=(\rho/m)^{1/2} e^{i S}$. The subscripts s and h refer to the soliton and halo, respectively, and $\whin$ represents the thermal velocity in the inner halo. Throughout this work, we loosely define the inner halo as the region surrounding the soliton ($r \gtrsim 6$--$8\,\rs$, where $\rs$ is the soliton half-density radius), characterized by approximately constant velocity dispersion, and the outer halo as the region near the halo virial radius, where the velocity dispersion declines (see the \textit{Velocity distribution} section). The angle brackets denote the mass-weighted root-mean-square velocity, for example, $\ave{w} \equiv \left( \int |\psi|^2 w^2 d^3x/\int |\psi|^2 d^3x \right)^{1/2}$.

\eref{eq:Ms} can be derived directly from the soliton scale transformation, where $\Ms$ is the soliton mass within $\rs$. Assuming virialization, $\Ep = -2\Ek = -\Mh(\avewh^2 + \avevh^2)$, where $\Ek$ is the sum of thermal and kinetic energy. $\alpha^2$ represents the fraction of $\Ek$ in thermal energy, which is $1/2$ assuming energy equipartition. $\beta$ describes the nonisothermality of temperature distribution within the halo and equals unity for an isothermal distribution. $\gamma$ characterizes the temperature contrast between the soliton and inner halo, with $\gamma=1$ representing thermal equilibrium. In the following sections, we demonstrate that $\alpha^2=1/2$ and $\gamma = 0.89$ provide reasonable approximations (Figs. \ref{fig:velocity} and \ref{fig:ratios}), and that both $\Ep(\Mh,c)$ and $\beta(c)$ can be inferred from an FDM concentration-mass-nonisothermality relation (\fref{fig:conc_and_temp} and \eref{eq:temp_c}). We also validate halo virialization and soliton fidelity (Fig. \ref{fig:ratios}), which are implicitly assumed in Eqs. (\ref{eq:Ms}) and (\ref{eq:Ws}).

Before examining in detail the contributions of individual terms in Eqs. (\ref{eq:Ms}) and (\ref{eq:Ws}), we present our key results in \fref{fig:sim_vs_theory} by comparing $\Ms$ from cosmological simulations with theoretical predictions. It demonstrates that both the model of \cite{schive2014b} and our refined model align well with the simulation results, with our model exhibiting a $1\sigma$ deviation of less than $30\%$ across the parameter space probed.


\textit{Numerical convergence}.---We first validate numerical convergence since it is critical for establishing a robust SHR. \fref{fig:profile} illustrates this by comparing fiducial- and low-resolution simulations, utilizing roughly 10 and 5 cells per granule size, respectively. The central profiles of both cases match the soliton solution. However, outside the soliton, only the fiducial run aligns well with not only the NFW model but also the N-body result \cite{veltmaat2018, Chiu2025, Chan2025}, whereas the low-resolution run generates an overly concentrated halo and a soliton that is $30\%$ more massive. See also \emph{Supplemental Material} \cite{SM}.

The large discrepancy in the low-resolution run stems from an underestimation of quantum pressure, leading to unphysical halo contraction. This contraction deepens the gravitational potential and increases the halo velocity dispersion, thereby raising the soliton energy and mass. Crucially, increasing resolution only within the soliton does not eliminate this problem, as the soliton properties depend on its thermal equilibrium with the surrounding halo granules (see \fref{fig:velocity}). Properly resolving the entire halo is thus essential for obtaining the correct SHR. This explains, to some extent, the overgrowth of soliton mass in the simulated massive halos at lower redshifts in \cite{schive2014b}. It may also partially account for the significant scatter and the different SHR reported in previous studies, particularly for massive halos (or larger $\mFDM$) with smaller granules.


\textit{Velocity distribution}.---After confirming numerical accuracy, we proceed to scrutinize the SHR. \fref{fig:velocity} presents the spherically averaged profiles of the thermal velocity $w$ and bulk velocity $v$. It shows that the average thermal velocity of the soliton $\avews$ within $r \lesssim 3.3 \rs$ approximately coincides with the inner-halo thermal velocity $\whin$ within $6\rs \lesssim r \lesssim 40\rs$. This result demonstrates that the soliton and inner halo are in thermal equilibrium (i.e., $\gamma \sim 1$ in \eref{eq:Ws}), a key assumption in the theoretical SHR. 

The soliton exhibits $w \gg v$ as it is supported by quantum pressure. In contrast, $w \sim v$ outside the soliton, signifying energy equipartition (i.e., $\alpha \sim 2^{-1/2}$ in \eref{eq:Ws}) \cite{mocz2017, Chowdhury2021}. Moreover, both $w$ and $v$ remain approximately constant in the inner halo but decrease in the outer halo, suggesting that only the inner halo follows an isothermal distribution. These findings challenge some assumptions in \cite{schive2014b}. First, their model neglects energy equipartition, leading to an overestimation of the halo thermal velocity. Second, they assume a spherical top-hat collapse model \footnote{An improved model considering the NFW profile is provided by \cite{bar2018}.} without accounting for the non-isothermal distribution, thus underestimating the inner-halo velocity. These two errors roughly cancel each other out, which explains why their simplified SHR still fits the simulation data (as shown in \fref{fig:sim_vs_theory}). In comparison, our improved model explicitly incorporates both factors, offering the possibility of exploring halos with larger temperature contrasts.

In addition, \fref{fig:velocity} shows that the inner-halo velocity is consistent with the velocity dispersion derived from the isotropic spherical Jeans equation (see \emph{Supplemental Material} \cite{SM}). This validation aligns with the findings of \cite{Chowdhury2021} and reinforces our results.


\textit{Concentration-mass-nonisothermality relation}.---To further investigate the non-isothermal velocity distribution revealed in \fref{fig:velocity}, we plot in \fref{fig:conc_and_temp} the relations between halo concentration $c$, halo mass $\Mh$, and nonisothermality $\beta=\whin/\avewh$ (see \eref{eq:Ws}) for all halos. We compute $c$ by fitting the halo density profiles at $r > 6\rs$ (to exclude the soliton) with the NFW model. Panel (a) shows the $c$-$\Mh$ relation at $z=0$. For CDM, $c$ increases monotonically with decreasing $\Mh$, where we adopt the $c$-$\Mh$ relation of \cite{Ishiyama2021} computed by the tool \texttt{Colossus} \cite{Diemer2018}. In comparison, the FDM halo concentration decreases below a redshift-independent half-mode halo mass, $\Mhalf=3.8\times10^{10}\mFDM^{-4/3}\Msun$ \cite{Schive2016}. This decrease is caused by quantum pressure delaying the onset of halo formation below $\Mhalf$, thereby suppressing the halo central density. Note that the $c$-$\Mh$ relation is $z$-dependent while here we only plot the results at $z=0$ as an illustration. For a given $\Mh$ and $c(\Mh,z,\mFDM)$, we can infer $\Ep$ in \eref{eq:Ws}.

We also show the results of N-body simulations using the same FDM initial conditions for comparison. Concentration parameters in genuine FDM simulations are slightly but systematically lower than their N-body counterparts and the theoretical predictions of \cite{laroche2022,kawai2024}, especially for $\mFDM=0.8$. This discrepancy is mainly due to the presence of massive solitons that suppress density near the soliton-halo transition (e.g., see \fref{fig:profile} at $3\kpc \lesssim r \lesssim 6\kpc$) and small periodic boxes, causing an additional $\sim 8\%$ deviation in the predicted $\Ms$.

Panel (b) of \fref{fig:conc_and_temp} shows the $\beta$-$c$ relation, revealing a positive correlation between the two. This occurs because FDM halos are approximately isothermal at formation, and subsequent accretion of lower-density mass in the outskirts leads to an increase in $c$ and a more non-isothermal velocity distribution. The FDM nonisothermality-concentration relation can be fitted by
\be
\whin/\avewh = 0.281 \log c + 1.05,
\label{eq:temp_c}
\ee
insensitive to redshift.

Note that most halos in our simulations have $\Mh \lesssim \Mhalf$, where quantum effects are important, resulting in systematically lower concentration parameters for FDM halos compared to CDM. Therefore, whether \eref{eq:temp_c} can be applied to halos with $\Mh \gg \Mhalf$ remains an open question. This uncertainty cannot be addressed by the simplified SHR of \cite{schive2014b} and underscores the importance of isolating $\beta(c)$ from other terms in \eref{eq:Ws}.


\textit{Further validation}.---Figs. \ref{fig:profile} and \ref{fig:velocity} show only a single representative halo. Additionally, we have implicitly assumed virial condition and soliton fidelity in Eqs. (\ref{eq:Ms}) and (\ref{eq:Ws}). To provide more robust evidence, in \fref{fig:ratios} we validate these implicit assumptions and examine the contributions of individual terms in Eqs. (\ref{eq:Ms}) and (\ref{eq:Ws}) for all FDM halos in our simulations with $2\Ek/|\Ep| < 1.35$ \cite{Neto2007}. The results confirm that halos approach virialization (panel (a)) and energy equipartition (panel (b)) as $z \rightarrow 0$. Solitons and inner halos are approximately in thermal equilibrium (panel (c)), with $\avews/\whin$ scattering around an average of $0.89$. The ratio being slightly below unity warrants deeper investigation. Panel (d) confirms that the simulated central density profiles closely match the analytical soliton solution, especially when the soliton-halo systems become more stable as $z \rightarrow 0$.

Based on these findings, we implicitly assume $2\Ek/|\Ep|=M_{\rm s, sim}/M_{{\rm s}, \avews}=1$ in \eref{eq:Ws} and further adopt $\avewh/\avevh=1$ and $\avews/\whin=0.89$ in our SHR plotted in \fref{fig:sim_vs_theory}. These assumptions hold well at $z=0$ but introduce modest uncertainties at higher redshifts, highlighting the importance of separating individual terms in \eref{eq:Ws} to facilitate further refinement of SHR. Nevertheless, note that the fiducial values shown in all four panels of \fref{fig:ratios} are positively correlated with the inferred soliton mass. As a result, deviations between theory and simulations arising from the systematic trends of $2\Ek/|\Ep|>1$, $\avewh/\avevh<1$, and $M_{\rm s, sim}/M_{{\rm s}, \avews}<1$ at $z>0$ tend to counterbalance each other, thereby extending the applicability of our SHR to higher redshifts.

\textit{Conclusions}.---We propose a refined SHR for FDM (Eqs. (\ref{eq:Ms}) and (\ref{eq:Ws})) supported by an extensive set of cosmological simulations. Our results demonstrate that all virialized halos exhibit a stable soliton core, followed by an NFW halo that, with sufficient numerical resolution, closely matches the collisionless N-body simulations from identical initial conditions (\fref{fig:profile}). Solitons and surrounding halo granules are in thermal equilibrium  (Figs. \ref{fig:velocity} and \ref{fig:ratios}). The energy distribution of halos satisfies equipartition and exhibits a non-isothermal profile (Figs. \ref{fig:velocity} and \ref{fig:ratios}). The FDM halo concentration parameter decreases below a characteristic halo mass, distinctly different from CDM and aligning with theoretical predictions (\fref{fig:conc_and_temp}). Furthermore, we find a positive correlation between FDM halo concentration and nonisothermality (\fref{fig:conc_and_temp} and \eref{eq:temp_c}). Our work reaffirms the simplified SHR proposed by \cite{schive2014b} (\fref{fig:sim_vs_theory}), while providing a more comprehensive description that enables exploration of a broad range of FDM particle masses, halo masses, and redshifts, especially for halos with larger temperature contrasts.

\begin{acknowledgments}
\textit{Acknowledgments}.---
We thank Chun-Yen Chen for his significant contributions to the development of \texttt{GAMER}. We use \textsc{yt} \citep{yt} for data visualization and analysis. This research is partially supported by the National Science and Technology Council (NSTC) of Taiwan under Grant No. NSTC 111-2628-M-002-005-MY4 and the NTU Academic Research-Career Development Project under Grant No. NTU-CDP-113L7729. We thank to National Center for High-performance Computing (NCHC) for providing computational and storage resources.
\end{acknowledgments}

\textit{Data availability}.--- The simulation code \texttt{GAMER} is accessible at \href{https://github.com/gamer-project/gamer}{https://github.com/gamer-project/gamer}. A Python script for computing the theoretical SHR is available at \href{https://github.com/calab-ntu/fdm-soliton-halo-relation}{https://github.com/calab-ntu/fdm-soliton-halo-relation}.

\bibliographystyle{apsrev4-2}

\newcommand {\apjl}     {Astrophys. J. Lett. }
\newcommand {\apjs}     {Astrophys. J. Suppl. }
\newcommand {\mnras}    {Mon. Not. R. Astron. Soc. }
\newcommand {\aap}      {Astron. Astrophys. }
\newcommand {\jcap}     {J. Cosmol. Astropart. Phys. }
\newcommand {\na}       {New Astron. }
\newcommand {\araa}     {Annu. Rev. Astron. Astrophys. }
\newcommand {\physrep}  {Phys. Rep. }

\bibliography{ref}

\begin{CJK*}{UTF8}{bkai}
\include{supplement}
\end{CJK*}

\end{document}

%% file: supplement.tex
\appendix
\counterwithin{figure}{section}
\counterwithin{table}{section}
\counterwithin{equation}{section}
\renewcommand{\thefigure}{S\arabic{figure}}
\setcounter{figure}{0}
\renewcommand{\theequation}{S\arabic{equation}}
\setcounter{equation}{0}
\renewcommand{\thetable}{S\arabic{table}}
\setcounter{table}{0}
\setcounter{page}{1}

\title{Deciphering the Soliton-Halo Relation in Fuzzy Dark Matter \\  \textit{Supplemental Material}}

\author{Pin-Yu Liao (廖品瑜)\orcidlink{0009-0007-8469-5880}}
\affiliation{Institute of Astrophysics, National Taiwan University, Taipei 10617, Taiwan}
\affiliation{Department of Physics, National Taiwan University, Taipei 10617, Taiwan}

\author{Guan-Ming Su (蘇冠銘)\orcidlink{0009-0008-4889-9779}}
\affiliation{Department of Physics, National Taiwan University, Taipei 10617, Taiwan}

\author{Hsi-Yu Schive (薛熙于)\orcidlink{0000-0002-1249-279X}}
\email{hyschive@phys.ntu.edu.tw}
\affiliation{Institute of Astrophysics, National Taiwan University, Taipei 10617, Taiwan}
\affiliation{Department of Physics, National Taiwan University, Taipei 10617, Taiwan}
\affiliation{Center for Theoretical Physics, National Taiwan University, Taipei 10617, Taiwan}
\affiliation{Physics Division, National Center for Theoretical Sciences, Taipei 10617, Taiwan}

\author{Alexander Kunkel\orcidlink{0009-0007-2957-4277}}
\affiliation{Institute of Astrophysics, National Taiwan University, Taipei 10617, Taiwan}
\affiliation{Department of Physics, National Taiwan University, Taipei 10617, Taiwan}

\author{Hsinhao Huang (黃新豪)\orcidlink{/0000-0002-7368-1324}}
\affiliation{Institute of Astrophysics, National Taiwan University, Taipei 10617, Taiwan}
\affiliation{Department of Physics, National Taiwan University, Taipei 10617, Taiwan}

\author{Tzihong Chiueh (闕志鴻)\orcidlink{0000-0003-2654-8763}}
\affiliation{Institute of Astrophysics, National Taiwan University, Taipei 10617, Taiwan}
\affiliation{Department of Physics, National Taiwan University, Taipei 10617, Taiwan}
\affiliation{Center for Theoretical Physics, National Taiwan University, Taipei 10617, Taiwan}

\maketitle

\section{Simulation Details}
\label{sup:sim-details}

\subsection{Methods}
\label{sup:sim-methods}

We use the \texttt{GAMER} \cite{GAMER} code for FDM simulations. It supports adaptive mesh refinement (AMR), which automatically and dynamically adjusts the spatial and temporal resolution to focus computational resources on the most demanding and scientifically relevant regions. The code adopts a hybrid OpenMP/MPI/GPU parallelization scheme, enabling efficient utilization of computing power on heterogeneous systems. Load balancing across multiple CPUs and GPUs is achieved using a Hilbert space-filling curve. In addition to FDM, the code also supports conventional hydrodynamic, magnetohydrodynamic, and N-body simulations.

For cosmological FDM simulations, \texttt{GAMER} adopts a novel hybrid algorithm \cite{Kunkel2024} that combines a wave scheme on small scales using finer grids and a fluid scheme on large scales using coarser grids. On small scales, the code solves the comoving Schr\"{o}dinger-Poisson equations:
\be
\left[ i a^2 \frac{\partial}{\partial t} + \frac{\hbar}{2m}\nabla^2 - \frac{m}{\hbar}\phi \right] \psi = 0,
\label{eq:schroedinger}
\ee
\be
\nabla^2 \phi = 4\pi G a (\rho - a^3\rho_{0}),
\label{eq:poisson}
\ee
where $a$ is the scale factor, $\hbar$ is the reduced Planck constant, $\phi$ is the gravitational potential, $G$ is the gravitational constant, $\rho=m|\psi|^2$ is the comoving mass density, and $\rho_0(t)$ is the cosmic mean density. We solve \eref{eq:schroedinger} using a local pseudospectral method based on Fourier continuations with Gram polynomials \cite{Lyon2010}, which provides significantly higher accuracy than conventional finite-difference methods.

On large scales, the code solves the comoving Hamilton-Jacobi-Madelung equations: 
\be
a^2\frac{m}{\hbar}\frac{\partial \rho}{\partial t} + \bm{\nabla} \cdot (\rho \bm{\nabla} S) = 0,
\label{eq:fluid_continuity}
\ee
\be
a^2\frac{m}{\hbar}\frac{\partial S}{\partial t} + \frac{1}{2}|\bm{\nabla} S|^2 + \frac{m^2}{\hbar^2} \phi
- \frac{1}{2} \frac{\nabla^2 \sqrt{\rho}}{\sqrt{\rho}} = 0,
\label{eq:fluid_phase}
\ee
where $S$ is the real phase field of the wave function: $\psi = (\rho/m)^{1/2}e^{iS}$. The bulk velocity is inferred from $\bm{v}=(\hbar/m)\bm{\nabla} S$. We evolve \eref{eq:fluid_continuity} using a Monotonic Upstream-centered Scheme for Conservation Laws (MUSCL) and solve \eref{eq:fluid_phase} using an upwind scheme, with third-order Runge-Kutta time integration.

This hybrid fluid-wave approach allows the use of much lower resolution to capture large-scale structure with the fluid scheme, which does not require resolving the short de Broglie wavelength associated with fast but smooth flows, while still resolving small-scale interference fringes and solitons with the wave scheme. However, there are a few caveats. First, the fluid formulation fails in regions of vanishing density (e.g., vortices), where the phase becomes discontinuous and the quantum pressure term $\nabla^2 \sqrt{\rho}/\sqrt{\rho}$ diverges. Second, the de Broglie wavelength in wave regions must be adequately resolved. To address these challenges, we have developed dedicated grid refinement criteria to ensure that the fluid scheme is applied only in smooth regions without strong interference, and that wave regions are resolved with sufficient resolution. Special care is taken at fluid-wave boundaries to ensure accurate conversion between the wave function $\psi$ and the fluid variables $(\rho, S)$ \cite{Kunkel2024}. Additionally, a density-based refinement criterion is used to adjust the resolution within halos when needed.

We perform a total of 31 simulations. \tref{tab:simulation_para} summarizes the simulation parameters. The initial conditions are generated at $z=100$ with periodic boundary conditions using \texttt{MUSIC} \cite{MUSIC}, where the input FDM linear power spectrum is computed by \texttt{AxionCAMB} \cite{AxionCAMB} with Planck cosmological parameters \cite{plank2018}. All simulations reach $z=0$. Initially, the entire simulation domain is evolved using the fluid scheme. As strong interference patterns gradually emerge, we apply local grid refinement and switch to the wave scheme in those regions. As a result, all halos and filaments are evolved using the wave scheme.

\begin{table}[t!]
\centering
\caption{Simulation parameters.}
\begin{tabular}[c]{ |c|c|c|c| } 
 \hline
 $\mFDM$ & Box size (Mpc) & Highest resolution (kpc) & Runs \\
 \hline
 0.1 & 5.2 / 5.9 & 0.32 &  7 \\ 
 0.2 & 4.2       & 0.13 & 14 \\
 0.8 & 2.1 / 3.0 & 0.09 & 10 \\ 
 \hline
\end{tabular}
\label{tab:simulation_para}
\end{table}

We select 42 halos with no major mergers since $z = 2$. For each halo, we compute the spherically averaged density and velocity profiles within the halo virial radius $\rh$. The halo mass $\Mh$ is defined as the enclosed mass within $\rh$, corresponding to a mean density $\zeta(z)$ times higher than the cosmic mean density $\rho_{0}(z)$:
\be
\Mh = \frac{4\pi}{3} \rh^3 \zeta (z) \rho_{0}(z),
\label{eq:halo_mass}
\ee
where
\be
\zeta(z) = \frac{18\pi^2 + 82[\Omega_m(z)-1] - 39[\Omega_m(z)-1]^2} {\Omega_m(z)}
\label{eq:zeta}
\ee
and $\Omega_m$ is the matter density parameter \cite{BN1998}. The scale radius of the outer NFW profile of an FDM halo can be inferred from
\be
r_{\rm NFW} = \frac{\rh}{c_{\rm FDM}},
\label{eq:scale_radius}
\ee
where $c_{\rm FDM}$ is the FDM concentration parameter. In \fref{fig:conc_and_temp}, $c_{\rm FDM}$ of simulation data is determined by fitting the outer halo profiles, while the theoretical predictions are obtained from
\be
c_{\rm FDM}(\Mh,z,\mFDM)=c_{\rm CDM}(\Mh,z)F\left(\frac{\Mh}{\Mhalf(\mFDM)}\right),
\label{eq:c_FDM}
\ee
where the CDM concentration parameter $c_{\rm CDM}(\Mh,z)$ is computed using the tool \texttt{Colossus} \cite{Diemer2018}, and the suppression term $F$ is defined as $F(x)=(1+ax^b)^c$ with $(a, b, c) = (5.496, -1.648, -0.417)$ \cite{laroche2022,kawai2024}.

The soliton density profile can be well described by the redshift-independent fitting function \cite{schive2014a}
\be
\rho_{\rm s}(r) = \frac{1.95\times10^7\mFDM^{-2}(\rs/\kpc)^{-4}} {[1+9.06\times10^{-2}(r/\rs)^2]^8} \Msun\kpc^{-3},
\label{eq:soliton_profile}
\ee
where the soliton radius $\rs$ is defined as the radius where the density drops to half its peak value. The soliton mass $\Ms$ is defined as the enclosed mass within $\rs$, satisfying the relation
\be
\Ms = 5.4\times10^7 \mFDM^{-2} \left(\frac{\rs}{\kpc}\right)^{-1} \Msun.
\label{eq:soliton_mass_radius}
\ee

For a given halo density profile $\rho_{\rm h}(r)$ and assuming spherical symmetry, we can infer the halo velocity dispersion $\sigma$ by solving the spherical Jeans equation \cite{GalacticDynamics2008}:
\be
\frac{d(\rho_{\rm h} \sigma_{\rm r}^2)}{dr} + 2\frac{\mu (\rho_{\rm h}\sigma_{\rm r}^2)}{r} \\
= -\rho_{\rm h} \frac{d\phi}{dr},
\label{eq:jeans}
\ee
where $\mu=1-\sigma_{\rm t}^2/2\sigma_{\rm r}^2$ is the anisotropy parameter, and $\sigma_{\rm r}$ and $\sigma_{\rm t}$ denote the radial and tangential velocity dispersions, respectively. To close \eref{eq:jeans}, one must assume a functional form for $\mu(r)$. Given that the granular structure of FDM halos is approximately isotropic, we assume an isotropic dispersion with $\mu=0$. Moreover, since an FDM halo outside the soliton is supported by both random bulk motion ($v$) and quantum pressure associated with thermal velocity ($w$), the dispersion $\sigma$ obtained from the Jeans equation should be interpreted as an \emph{effective} velocity dispersion incorporating both contributions. Accordingly, we expect $3\sigma_{\rm r}^2 \sim v^2 + w^2 \sim 2v^2$, as confirmed in \fref{fig:velocity}.

\subsection{Example results}
\label{sup:sim-results}

\begin{figure*}[ht!]
\includegraphics[width=\textwidth]{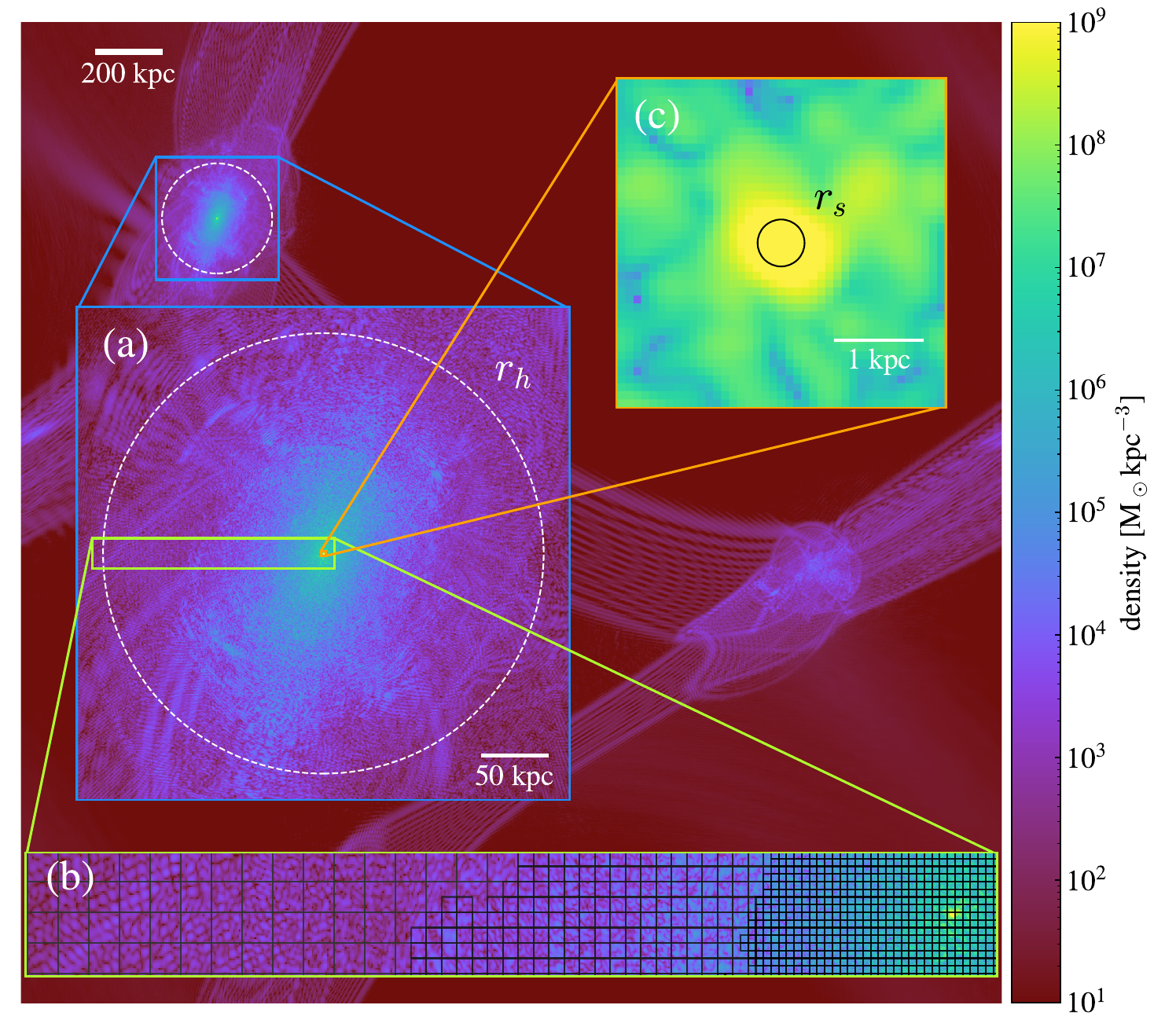}
\caption{
A slice of FDM density field on various scales through a $\Mh \sim 2.6\times10^{11}\Msun$ halo with $\mFDM=0.8$ at $z=0$. Dashed and solid circles indicate the halo virial radius $\rh$ and soliton radius $\rs$, respectively, with a corresponding soliton mass of $\Ms=3.0\times10^8\Msun$. The inset at the bottom illustrates the distribution of AMR grids at various radii, where each grid comprises $16\times16\times16$ cells in three dimensions. The maximum spatial resolution is $0.09\kpc$.
}
\label{fig:slice_zoom}
\end{figure*}

\fref{fig:slice_zoom} shows a slice of FDM density field with $\mFDM=0.8$ at $z=0$ in a periodic comoving box of length $L=3\Mpc$. The slice cuts through a $\Mh \sim 2.6\times10^{11}\Msun$ halo, highlighting FDM features across multiple scales. The image of the full simulation box reveals the filamentary structure characterized by transverse interference fringes. Inset (a) displays the entire halo, exhibiting ubiquitous density granules on the de Broglie scale arising from stochastic constructive and destructive interference. The dashed circle marks the virial radius $\rh$.

Inset (b) highlights the distribution of density granules inside a halo, with the AMR grid overlaid. The granule size decreases toward the halo center due to increasing velocity (see \fref{fig:velocity}). It demonstrates that our grid refinement scheme, achieving a maximum spatial resolution of $0.09\kpc$, can accurately resolve the de Broglie scale across all radii.

Inset (c) shows a close-up of the central soliton. The solid circle represents the soliton radius $\rs=0.27 \kpc$, corresponding to a soliton mass of $\Ms=3.0\times10^8\Msun$. The soliton-halo transition occurs at $\sim 3.3\rs$ \cite{mocz2017, Chiang2021}, which is well resolved by approximately ten cells.

\begin{figure*}[ht!]
\centering
\includegraphics[width=\textwidth]{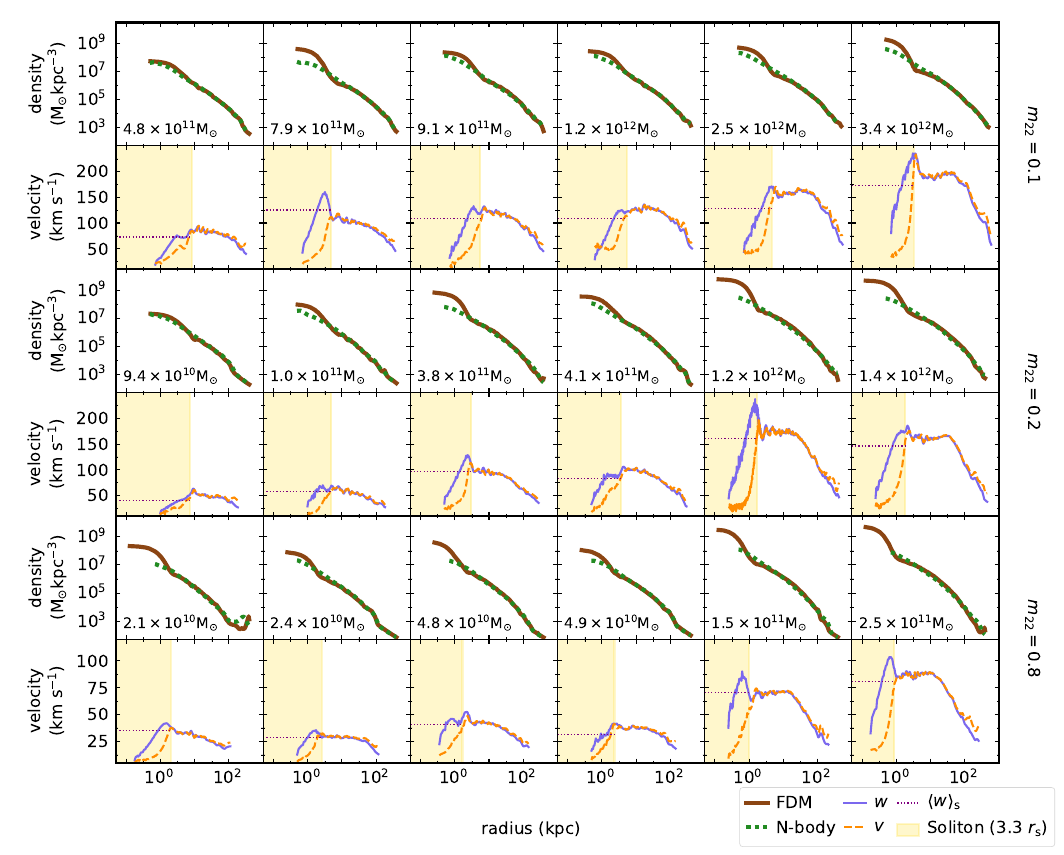}
\caption{
Density and velocity profiles of 18 halos with various $\mFDM$ and halo mass at $z=0$. The soliton regions are highlighted in yellow. In all cases, the FDM density profiles (thick solid lines) closely match the N-body simulations (thick dotted lines) outside the soliton, consistent with \fref{fig:profile} and demonstrating numerical convergence (see also \fref{fig:convergence}). The thin solid, thin dashed, and horizontal dotted lines represent the thermal velocity profiles $w$, bulk velocity profiles $v$, and average thermal velocity within the soliton $\avews$, respectively. All systems exhibit energy equipartition and an isothermal distribution in the inner halo surrounding the soliton, as well as thermal equilibrium between the soliton and inner halo, in agreement with \fref{fig:velocity}.
}
\label{fig:18halos}
\end{figure*}

\fref{fig:18halos} shows the density and velocity profiles of 18 halos at $z=0$, spanning a representative range of FDM particle mass and halo mass from our simulation set: $\mFDM=0.1$--$0.8$ and $\Mh \sim 2\times10^{10}$--$3\times10^{12}\Msun$. All halos exhibit a prominent, stable soliton core, followed by an NFW halo that aligns well with the collisionless N-body simulations from the same initial conditions. Consistent with \fref{fig:velocity}, the thermal and bulk velocities have similar magnitudes and remain roughly constant in the inner halo surrounding the soliton, demonstrating energy equipartition and an isothermal distribution. Moreover, the average thermal velocity of the soliton matches the inner-halo thermal velocity, confirming thermal equilibrium between the two.

\subsection{Convergence tests}
\label{sup:sim-convergence}

\begin{figure}[ht!]
\includegraphics[width=\columnwidth]{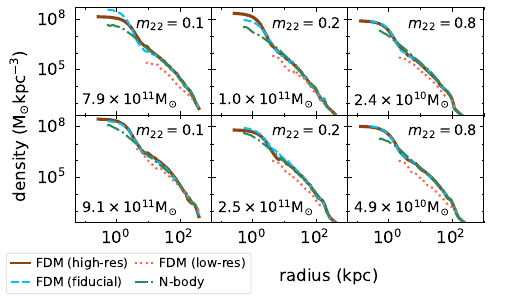}
\caption{
Convergence tests of the density profiles for six halos with different $\mFDM$ and halo mass at $z=0$. The high-resolution (solid lines) and fiducial-resolution (dashed lines) results show good agreement, with only minor discrepancies in the central regions due to stochastic soliton oscillations. Furthermore, outside the solitons, both profiles closely match the N-body simulations (dash-dotted lines). By contrast, the low-resolution results (dotted lines) exhibit clear deviations from the other three cases.
}
\label{fig:convergence}
\end{figure}

We perform convergence tests to validate the soliton-halo relation by increasing resolution until each FDM halo satisfies the following criteria: (i) the central density profile matches the analytical soliton solution \cite{schive2014b}, and (ii) the halo profile outside the central soliton agrees with the N-body simulation \cite{veltmaat2018, Chiu2025, Chan2025}. This necessitates using at least 6--12 cells per halo granule size, defined as $d \sim 0.35 \LdB$, where $\LdB \sim 1.48 ( v / 100\kms )^{-1} \mFDM^{-1} \kpc$ and $\bm{v}$ is the three-dimensional bulk velocity \cite{BarOr2019}. The corresponding maximum spatial resolutions are $0.32$, $0.13$, and $0.09 \kpc$ for $\mFDM=0.1, 0.2$ and $0.8$, respectively. For several representative halos, we additionally verify that (iii) the entire halo profile converges at even higher resolution.

\fref{fig:convergence} shows the convergence tests for six halos with different $\mFDM$ and halo mass at $z=0$, focusing on criteria (ii) and (iii). The fiducial runs resolve each density granule with approximately 10 cells. The high-resolution cases double the resolution, while the low-resolution cases reduce it by a factor of four. Outside the solitons, the high-resolution, fiducial-resolution, and N-body results align well, thereby fulfilling criteria (ii) and (iii). By contrast, the low-resolution profiles are systematically lower, primarily due to under-resolving the de Broglie wavelength associated with high-speed inflows, which delays mass accretion.

Our results show that criterion (iii) is always met when criteria (i) and (ii) are satisfied. While criterion (i) is relatively easy to achieve, criteria (ii) and (iii) are more challenging. Furthermore, we emphasize that matching an NFW profile outside the soliton does not guarantee the satisfaction of criterion (ii), as an inaccurate profile resulting from insufficient resolution may still match an \emph{incorrect} NFW profile. Since high-resolution FDM simulations are extremely expensive, conducting N-body counterpart simulations proves both effective and efficient for confirming the convergence of FDM simulations (see also \fref{fig:18halos}).